\documentclass{aa}
\usepackage{amsmath,graphicx,amssymb}
\usepackage[varg]{txfonts}
\usepackage{natbib}
\usepackage{pst-eps}
\bibpunct{(}{)}{,}{a}{}{,}

\newcommand{\zav}[1]{\left(#1\right)}

\newlength\staretab

\newcommand\de{\text{d}}

\newcommand\hp{\ensuremath{H_\text{p}}}

\newcommand\msr{\ensuremath{M_\odot\,\text{yr}^{-1}}}

\newcommand\intvidpo{\!\!\int\limits_{\begin{array}{c}\text{\scriptsize
visible}\\[-2mm]\text{\scriptsize surface}\end{array}}\!\!}

\begin{document}

\title{Light variations due to the line-driven wind instability and wind
blanketing in O stars}

\author{J.~Krti\v{c}ka\inst{1} \and A.~Feldmeier\inst{2}}


\institute{\'Ustav teoretick\'e fyziky a astrofyziky, Masarykova univerzita,
           Kotl\'a\v rsk\' a 2, CZ-611\,37 Brno, Czech Republic 
           \and
           Institut f\"ur Physik und Astronomie, Universit\"at Potsdam,
           Karl-Liebknecht-Stra{\ss}e 24/25, 14476 Potsdam-Golm, Germany}

\date{Received}

\abstract{A small fraction of the radiative flux emitted by hot stars is
absorbed by their winds and redistributed towards longer wavelengths. This
effect, which leads also to the heating of the stellar photosphere, is termed
wind blanketing. For stars with variable winds, the effect of wind blanketing
may lead to the photometric variability. We have studied the consequences of line
driven wind instability and wind blanketing for the light variability of O
stars. We combined the results of wind hydrodynamic simulations and of global
wind models to predict the light variability of hot stars due to the wind
blanketing and instability. The wind instability causes stochastic light
variability with amplitude of the order of tens of millimagnitudes and a typical
timescale of the order of hours for spatially coherent wind structure. The
amplitude is of the order of millimagnitudes when assuming that the wind
consists of large number of independent concentric cones. The variability with
such amplitude is observable using present space borne photometers. We show that
the simulated light curve is similar to the light curves of O stars obtained
using BRITE and CoRoT satellites.}


\titlerunning{Light variations due to the line-driven wind instability and wind
blanketing in O stars}

\authorrunning{J.~Krti\v{c}ka, A.~Feldmeier}
\maketitle

\section{Introduction}

The advent of photometers on board dedicated spacecrafts like Kepler, 
CoRoT, MOST, and BRITE revolutionized the field of astronomical photometry. The
space-borne photometers provide not only light curves with unprecedented
coverage, but they also lead to a significant improvement of the precision of
measurements. As a result, stars that were deemed constant were found to be
variable leading to discoveries of new types of variable stars.

Single O stars are a typical examples of stars for which no variability was
expected, yet that still show low-amplitude light variations. For example, significant
low-frequency variability was detected in Kepler photometry of a blue supergiant
HD 188209 and was attributed to the gravity waves \citep{aeovar}. A dedicated
CoRoT observing run focused on the young open cluster NGC 2244 detected
stochastic variations in O stars, which were attributed to the red noise
that is possibly due to the subsurface convection \citep{blomcor} and multiple
frequencies due to stellar oscillations \citep{decor,bricor,macor}. Similar
variations were also found after the subtraction of binary light curve in
$\delta$~Ori~Aa \citep{mostdelori}, in early-B supergiant HD 2905
\citep{simondia}, and in $\zeta$~Pup \citep{ram}.

Low-amplitude light variability in O stars may be connected to the line-driven
wind instability. As a result of the Doppler effect, wind driving by multiple
lines in hot stars is unstable \citep{lusol,ornest}. The numerical simulations
have shown that the line-driven wind instability leads to the variability of
the wind density, velocity, and mass-loss rate on a typical scale of hours
\citep{ocr,felpulpal}. Due to supersonic nature of hot star winds, the wind
instability leads also to appearance of wind shocks  and the wind X-ray emission
\citep{felpulpal}. The line-driven wind instability is likely connected with
clumping, which affects observed wind spectral features
\citep[e.g.][]{chuchcar,sund,clres1,clres2,shendelori} and therefore affects the
wind mass-loss rate estimates.

Wind instabilities are self-initiating \citep{ocr}, but may be also initiated
and modulated by photospheric motions \citep{felpulpal}. Consequently, the
instabilities may be connected with photospheric turbulence whose presence
affects, for example, the widths of photospheric lines \citep{ae,cant,jian}. 

Part of the flux emerging from the photospheres of stars is absorbed by the wind
and emitted back to the stellar photosphere. The backwards emission causes
heating of the photosphere especially at low optical depths and also leads to
redistribution of the flux mainly from the short-wavelength part of spectrum to
longer wavelengths. This effect is termed wind blanketing \citep{acko} and is
important for obtaining precise effective temperatures of hot stars with strong
winds \citep{boh,prvnifosfor,okali}. The wind blanketing redistributes the
emergent flux in dependence of wind opacity, and amount of redistribution
depends on the wind mass-loss rate. Therefore, wind blanketing does not lead to
any light variability for fixed mass-loss rate.

On the other hand, if the mass-loss rate is variable, then the wind blanketing
is modulated by the mass-loss rate, resulting in the photometric variability.
For example, in magnetic hot stars the wind mass flux depends on the tilt of the
magnetic field \citep{owoudan}. Therefore, the strength of the wind blanketing
varies across the stellar surface, which, due to stellar rotation, leads to the
photometric variability of magnetic O stars. This effect can partly explain the
photometric light curve of HD~191612 \citep{magvar}.

The line-driven wind instability is another process that leads to the mass-loss
rate variation \citep{ocr,felpulpal,runow}. Therefore, also the line-driven wind
instability causes photometric variability in hot stars. This can be connected
with the low-amplitude variability observed in O stars. To study this
possibility, we used hydrodynamical simulation of \citet{felpulpal} to derive the
mass-loss rate variation and the relation between the mass-loss rate and
photometric flux \citep{magvar} to predict the light variability.

\section{Modelling of the light variability: spherically symmetric case}

The calculation of the light variability of hot stars due to the line-driven
wind instability is a very complex problem. In general, 3D time-dependent
hydrodynamical simulations of the stellar wind are required to obtain the
density and velocity structure of the circumstellar environment of hot stars and
its variability. Such simulations have to be coupled to a global (unified) 3D
wind model that includes the influence of the wind on the stellar
photosphere (accounting for wind blanketing) and that is able to solve the
radiative transfer equation for moving media whose ionization and excitation
state is out of equilibrium. There is no such code that would be able to treat
the whole complex problem. Therefore, we used existing wind simulations and
global wind models and combine their output to derive the expected light
variations in our stars. Due to the simplifications involved, we have not focussed on
a particular star or stellar type, but instead we aim to obtain general magnitude of the
effect and its properties.

\subsection{Hydrodynamical simulations}

The hydrodynamical simulations we have employed here were described by
\citet{felpulpal} in detail. The simulations solved continuity, momentum, and
energy equation for a spherically symmetric non-rotating wind flow. The
simulations were performed using the  smooth source function method
\citep{Owocki1,Owocki2}.  The hydrodynamical part of the code is based on a
standard van Leer solver. The simulations use staggered mesh, operator splitting
of advection and source terms, advection terms in a conservative form using van
Leer's (\citeyear{vanLeer}) monotonic derivative as an optimised compromise
between stability and accuracy, Richtmyer artificial viscosity, and
non-reflecting boundary conditions \citep{Hedstrom,Thompson1,Thompson2}. A
turbulent variation of the velocity at a level of roughly one third of the sound
speed was introduced as seed perturbation for unstable growth at the wind base.

The hydrodynamical simulations were performed for specific stellar parameters
corresponding to $\zeta$~Ori~A, that is the effective temperature
$T_\text{eff}=31\,500\,\text{K}$, radius $R_*=24\,R_\odot$, and mass
$M=34\,M_\odot$, and yielded the mean mass-loss rate $\dot M_0=3\times10^{-6}\,\msr$.
The simulations are robust against changes in stellar parameters. Consequently,
the relative variations of wind density derived for such specific parameters are
also applicable to other O stars.

\begin{figure}[t]
\centering
\resizebox{\hsize}{!}{\includegraphics{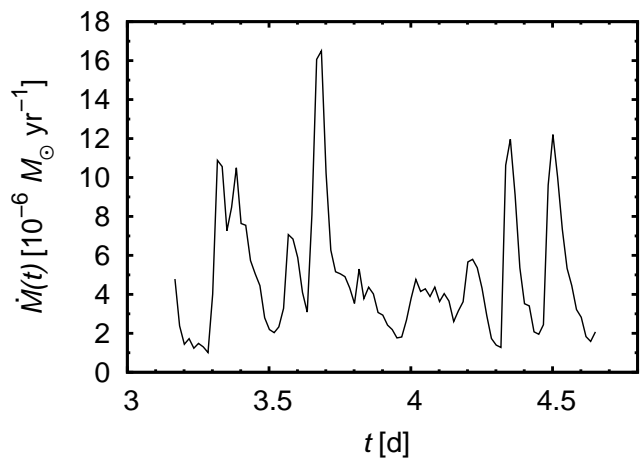}}
\caption{Mass-loss rate from the numerical simulations averaged over
$r\in[1.01\,R_*,1.02\,R_*]$ (see Eq.~\eqref{dmdtt})}
\label{d9ata_stormg_mast}
\end{figure}

The calculation of the wind blanketing directly from the hydrodynamic
simulations would require solution of radiative transfer and kinetic equilibrium
equations (also called NLTE equations) for non-monotonic velocity law. To our
knowledge, there is no such code that is able to treat this problem in its full
complexity. Therefore, to make the problem tractable, we calculate a mean value
of the mass-loss rate in the region close to the star, where the wind blanketing
originates, 
\begin{equation}
\label{dmdtt}
\dot M(t)=\frac{4\pi}{r_2-r_1}\int_{r_1}^{r_2}r^2\rho\left|v\right|\,\de r,
\end{equation}
where $\rho$ and $v$ are the wind density and radial velocity in particular
time. Most of the flux in the optical domain is emitted from the region below
the sonic point, which typically appears at radii $1.01-1.02\,R_*$ in the
numerical simulations. Therefore, we selected $r_1=1.01\,R_*$ and
$r_2=1.02\,R_*$. However, the final light curve does not significantly depend on
the choice of $r_1$ and $r_2$. As a result of this choice, in Eq.~\eqref{dmdtt}
we average over 50 grid points of hydrodynamical simulation. A plot of $\dot
M(t)$ is given in Fig.~\ref{d9ata_stormg_mast}.

\subsection{Global wind models}

The variations of the emergent flux with wind mass-loss rate were derived from
spherically symmetric stationary METUJE global wind models \citep{cmf2}. The
code calculates the wind model in a global (unified) approach that integrates
the description of the hydrostatic photosphere and supersonic wind. The model
solves the radiative transfer equation in the co-moving frame (CMF) with
opacities and emissivities calculated using occupation numbers derived from the
kinetic equilibrium (NLTE) equations. For given stellar parameters, the model
enables us to predict the radial dependence of the density, velocity, and
temperature from hydrodynamical equations and to derive the wind mass-loss rate
$\dot M$ and the stellar emergent flux that accounts for the wind blanketing.

For our modelling we assumed a typical parameters of O stars corresponding to
HD~191612, $T_\text{eff}=36\,000\,\text{K}$, $R_*=14.1\,R_\odot$, and
$M=29.2\,M_\odot$. To model the dependence of the emergent flux on the mass-loss
rate with fixed stellar parameters, we artificially scaled the radiative force
in the wind. This yields a series of wind models and emergent fluxes
$F(\lambda,\dot M)$ parameterized by the wind mass-loss rate. Because the amount
of the flux redistributed by the wind depends on the effective temperature, we
expect slightly different dependence for stars with different temperature.
However, because we are concerned with typical light variations, we limited our
study to just one star. On the other hand, such selection affects only the
amplitude of the light curve. The preliminary results derived for
300-1 model of \citet[the model has parameters close to that used in
hydrodynamical simulations]{cmf2} indicate that the amplitude of the
variability due to the wind blanketing does not significantly depend on stellar
parameters.

The light variability is derived from a relation between the $H_\text{p}$
magnitudes and mass-loss rate \citep[Eq.~(2)]{magvar}
\begin{equation}
\label{hpt}
\Delta H_\text{p}(t)=-2.5\log(e) \frac{\Delta F}{F_0}
\log\zav{\frac{\dot M(t)}{\dot M_0}},
\end{equation}
where $\Delta F=1.6\times10^7\,\text{erg}\,\text{s}^{-1}\,\text{cm}^{-2}$,
$F_0=9.0\times10^8\,\text{erg}\,\text{s}^{-1}\,\text{cm}^{-2}$
is the emergent flux for a reference mass-loss rate, and $\dot M_0$ is the mean
mass-loss rate. The relation was derived for HD~191612 parameters using METUJE
global wind models \citep{cmf2}. 

\subsection{Calculated light curve}

\begin{figure}[t]
\centering
\resizebox{\hsize}{!}{\includegraphics{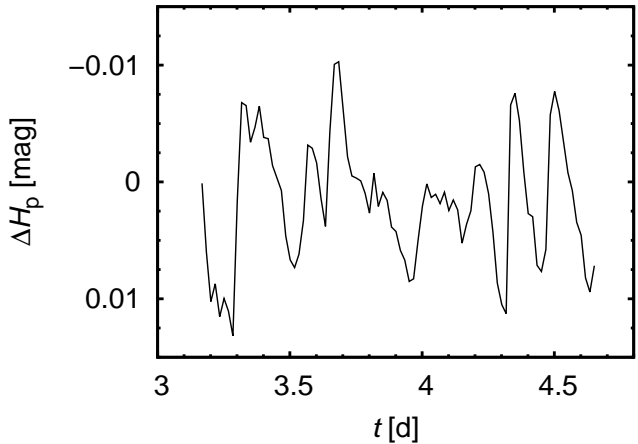}}
\caption{Light curve calculated using Eq.~\eqref{hpt} for the
mass-loss rate variations plotted in Fig.~\ref{d9ata_stormg_mast}}
\label{d9ata_stormg_var}
\end{figure}

The light curve calculated using Eq.~\eqref{hpt} for the mass-loss rate
variations plotted in Fig.~\ref{d9ata_stormg_mast} is given in
Fig.~\ref{d9ata_stormg_var}. The amplitude of the light curve is of the order
of 0.01\,mag and the typical timescale of the light variability (hours) is the
same as the timescale of the variability of the mass-loss rates as derived from
the simulations.

The amplitude of the light curve is proportional to the mass-loss rate
variations, which depend on the surface perturbations. The light variations are
therefore stronger for stronger base perturbations. The wind structure is
triggered by the photospheric perturbations, which therefore determine the time
scale of the variability.

\section{The case without spherical symmetry}

Within a spherical symmetry we assumed that mass-loss rate variations are
coherent across the stellar surface. This assumption is highly unrealistic for
real wind variations. However, the coherence length in the
horizontal direction is unclear. A typical patch size of the instability generated
structure is of the order of degrees, as inferred from observations
\citep[e.g.][]{desow} and the typical number of clumps in the observed part of
the wind is of the order of $10^3$--$10^5$ \citep{davo,osporcar,nazog,clres2}.

To account for the deviation from spherical symmetry, we followed the approach
of \citet{desow} and \citet{lidarikala} and assumed that the stellar wind
consists of $N$ concentric cones. We assumed that each of these cones is
independent, and that the mass-loss rate from the $i$-th cone with stellar
surface cross-section $\Omega_i R_*^2$ at time $t$ is 
\begin{equation}
\label{mlosconka}
\dot M_i(t)=\frac{\Omega_i}{4\pi}\dot M(t+\Delta t_i),
\end{equation}
where $\Delta t_i$ is random for each $i$ and $\dot M(t)$ is given by
Eq.~\eqref{dmdtt}. Moreover, we assumed that $\dot M(t)$ varies periodically
with the period given by the time extent of available simulations 
(which is about 1.5\,d, see Fig.~\ref{d9ata_stormg_mast}).

The cones were selected in such a way that $\Omega_i$ is roughly the same for
all cones. We divided the visible stellar surface into elements bounded by a
specified number of concentric rings. The first element that directly faces the
observer is assumed to specify just one cone. The number of cones specified by
other rings is selected in such a way that solid angle set by the cones is
roughly equal to solid angle of the first cone. The number of cones is
therefore a function of the number of rings. Because we specify the number
of latitudinal rings, we cannot calculate the light curves for an arbitrary
number of cones.

The magnitude difference between the observed flux $f_{\hp}(t)$ at a given time
and the reference flux $f_{\hp}^\text{ref}$ in passband $\hp$ is defined as
\begin{equation}
\label{velik}
\Delta \hp(t)=-2.5\,\log\,\zav{\frac{{f_{\hp}(t)}}{f_{\hp}^\mathrm{ref}}}.
\end{equation}
The reference flux is obtained under the condition that the mean magnitude
difference over the simulated lightcurve is zero. The radiative flux in a
passband $\hp$ at the distance $D$ from the spherical star is
\citep{hubenymihalas}
\begin{equation}
\label{vyptok}
f_{\hp}(t)=\zav{\frac{R_*}{D}}^2\intvidpo
I_{\hp}(\theta,\Omega,t)\cos\theta\,\text{d}\Omega.
\end{equation}
The intensity $I_{\hp}(\theta,\Omega,t)$, at angle $\theta$ with respect to the
normal to the surface was obtained at each surface point with spherical
coordinates $\Omega$ from the emergent flux taking into account the limb
darkening $u_{\hp}(\theta)$
\begin{equation}
I_{\hp}(\theta,\Omega,t)=u_{\hp}(\theta)\,I_{\hp}(\theta=0,\dot
M(\Omega,t))=\frac{u_{\hp}(\theta)}{\langle u_{\hp} \rangle}
F\!_{\hp}(\dot M(\Omega,t)),
\end{equation}
with limb darkening coefficients from \citet{okraaj}. Here $\langle u_{\hp}
\rangle=2\pi\int_0^{\pi/2}u_{\hp}(\theta)\cos\theta \sin\theta\,\de\theta$.
The emergent flux $F\!_{\hp}(\dot M(\Omega,t))$ is obtained from Eq.~\eqref{hpt}
modified for fluxes, and the mass-loss rate $\dot M(\Omega,t)$ is given by
Eq.~\eqref{mlosconka} for a cone that corresponds to given coordinates $\Omega$.

\begin{figure}[t]
\centering
\resizebox{\hsize}{!}{\includegraphics{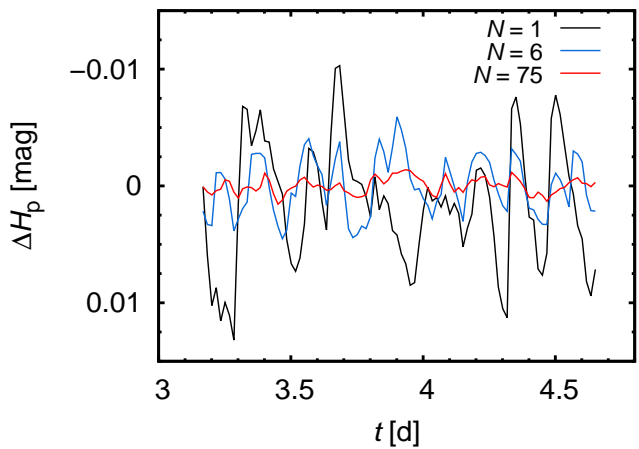}}
\caption{Light curve calculated using Eq.~\eqref{velik} for different number
$N$ of concentric cones}
\label{d9ata_stormg_varms}
\end{figure}

The resulting light curves are given in Fig.~\ref{d9ata_stormg_varms}. From this
figure it follows that the amplitude of the light variability due to wind
instability quickly decreases with the number of cones. On the other hand, even
with a relatively large number of cones, the light variability should be
detectable with satellite photometry.

\section{Comparison with observations}

\begin{figure}[t] 
\centering
\resizebox{\hsize}{!}{\includegraphics{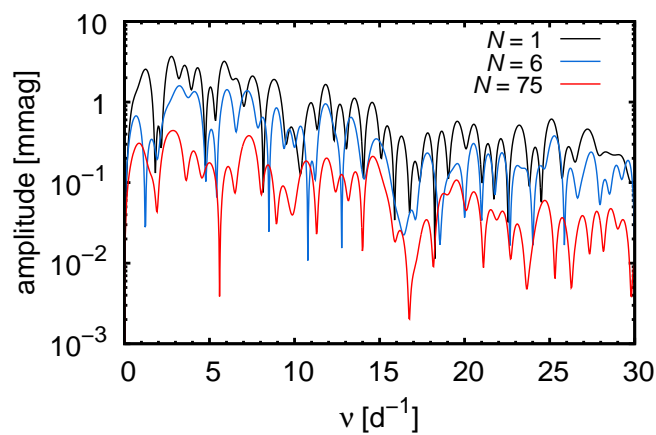}}
\caption{Fourier analysis of the light curves given in
Fig.~\ref{d9ata_stormg_varms}}
\label{d9ata_stormg_varmsf}
\end{figure}

We used Period04 \citep{period} to analyse the simulated light curves (see
Fig.~\ref{d9ata_stormg_varmsf}). The general shape of the Fourier spectrum does
not significantly depend on the number of cones used to calculate the light
curve. The Fourier spectrum shows maximum at the frequency of about
$3\,\text{d}^{-1}$ which corresponds to the characteristic times scale of the
variability of the mass-loss rate. The location of the maximum peak is nearly
independent of the number of assumed cones. There are further peaks at higher
frequencies. An additional low-frequency peak at about $1\,\text{d}^{-1}$ is most
likely connected with the length of available simulations.

Due to a stochastic nature of the light curves, a direct comparison of the
simulated and observed light curves is not meaningful. On the other hand, it is
possible to compute the Fourier spectrum of the light curves, fit the
spectrum by some phenomenological model, and compare the fit parameters derived
for observed and simulated light curves. Similar techniques are used to study
time series in various contexts, for example, light curves of X-ray binaries
\citep{burder}, solar flare data \citep{thrl}, or to search for granulation in
Cepheids \citep{derek}.

To compare the theoretical light curves with observations, we fitted the
Fourier spectrum $A(\nu)$ in Fig.~\ref{d9ata_stormg_varmsf} by a polynomial
\begin{equation}
\label{complicated}
\log\zav{\frac{A(\nu)}{1\,\text{mmag}}}=a\zav{\frac{\nu}{1\,\text{d}^{-1}}}+b.
\end{equation}
The fit parameters given in Table~\ref{fittab} demonstrate that the shape of the
Fourier spectrum (given by $a$ parameter) does not depend on the assumed number
of cones, while the amplitude of peaks (described by $b$) depends on
$N$. The fit of results from Table~\ref{fittab} shows that the parameter
$b$ varies with $N$ on average as
\begin{equation}
\label{bvarn}
b=-0.45\log N-2.88.
\end{equation}

\begin{table}[t]
\caption{Parameters of the fit Eq.~\eqref{complicated} of the Fourier spectrum}
\label{fittab}
\centering
\begin{tabular}{llcc}
\hline
Parameter&&$a$&$b$\\
\hline
Simulation & $N=1$  & $-0.030$ & $-2.87$ \\
           & $N=6$  & $-0.027$ & $-3.25$ \\
           & $N=75$ & $-0.034$ & $-3.71$ \\
\hline
Star & HD 37128 & $-0.11$  & $-2.93$ \\
     & HD 46150 & $-0.035$ & $-4.26$ \\
     & HD 46223 & $-0.037$ & $-3.94$ \\
     & HD 46966 & $-0.032$ & $-4.49$ \\
\hline
\end{tabular}
\end{table}

\begin{figure}[t]
\centering
\resizebox{\hsize}{!}{\includegraphics{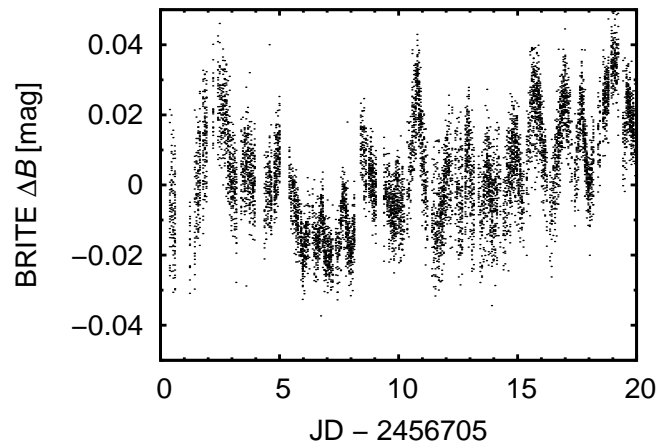}}
\caption{Portion of the light curve of $\epsilon$~Ori derived using the BRITE
satellite. Difference between the actual and mean magnitudes is plotted.}
\label{hd37128}
\end{figure}

\begin{figure}[t]
\centering
\resizebox{\hsize}{!}{\includegraphics{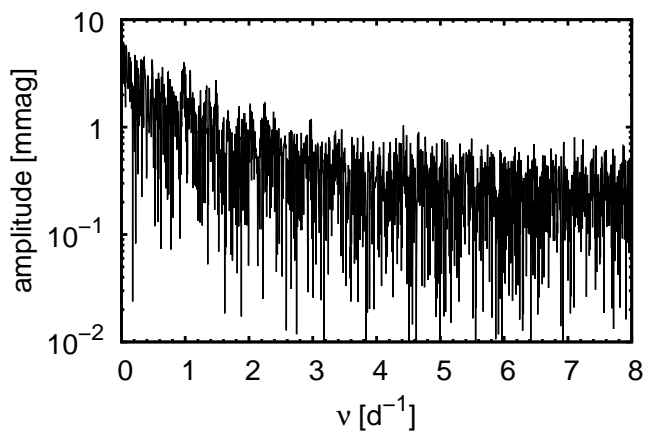}}
\caption{Fourier spectrum of the light curve of $\epsilon$~Ori}
\label{hd37128f}
\end{figure}

The simulated light curve can be compared, for example, with the light curve of
$\epsilon$~Ori (\object{HD 37128}, B0Ia) derived using the BRITE satellites. The
BRITE-Constellation consists of six satellites working in blue and red domains
of visible spectra \citep{brite}. The satellites provide high-precision
photometry of bright objects including chemically peculiar stars, pulsating
stars, and Be stars \citep{baadebrit,weissbrit,dadabrit,handbrit}. The data were
downloaded from the BRITE public data
archive\footnote{https://brite.camk.edu.pl/pub/index.html}. We used observations
of $\epsilon$~Ori obtained by BAb (BRITE-Austria) satellite between JD 2456628
-- 2456734. We used observations in a blue filter which covers the wavelength
range of 390 -- 460\,nm \citep{kamow}. The observed differential light curve
shows complex variability with amplitude comparable to the simulated light
variations in Fig.~\ref{d9ata_stormg_var} and a typical timescale of about 1\,d
(see Fig.~\ref{hd37128} and also \citealt{adueps}). A corresponding peak
appears also in the Fourier spectrum in Fig.~\ref{hd37128f}. Consequently, while
the time scale of the variability is comparable to that derived from
simulations, the interpretation of the amplitude of the observed variability
would require either a coherent wind variability or a very strong base
perturbation.

The inspection of the BRITE archive revealed that a similar type of variability
is also likely present in two additional stars $\zeta$~Ori~A (O9.2Ib, \object{HD
37742}, HR 1948) and J Pup (B0.5Ib, \object{HD 64760}, HR 3090). The light
variations of these stars have roughly the same amplitude and the time scale as
light variations of $\epsilon$~Ori \citep[see also][]{buybrite}. The star
$\zeta$~Ori~A has weak magnetic field and the rotational period of about 6.8\,d
\citep{blazar} and therefore it enables us to study the interaction of
atmospheric turbulent motions and magnetic field. Both stars show the line
profile variability due to corotating interacting regions \citep{howdac,kapdac},
which are likely unrelated to large-scale magnetic fields \citep{adu}.
Because the line-driven wind instability is possibly
present in all hot star winds being one of the sources of their
X-ray emission \citep[e.g.][]{igor,naze}, the light
variations due to to corotating interacting regions and wind instabilities
coexist.

\begin{figure}[t]
\centering
\resizebox{\hsize}{!}{\includegraphics{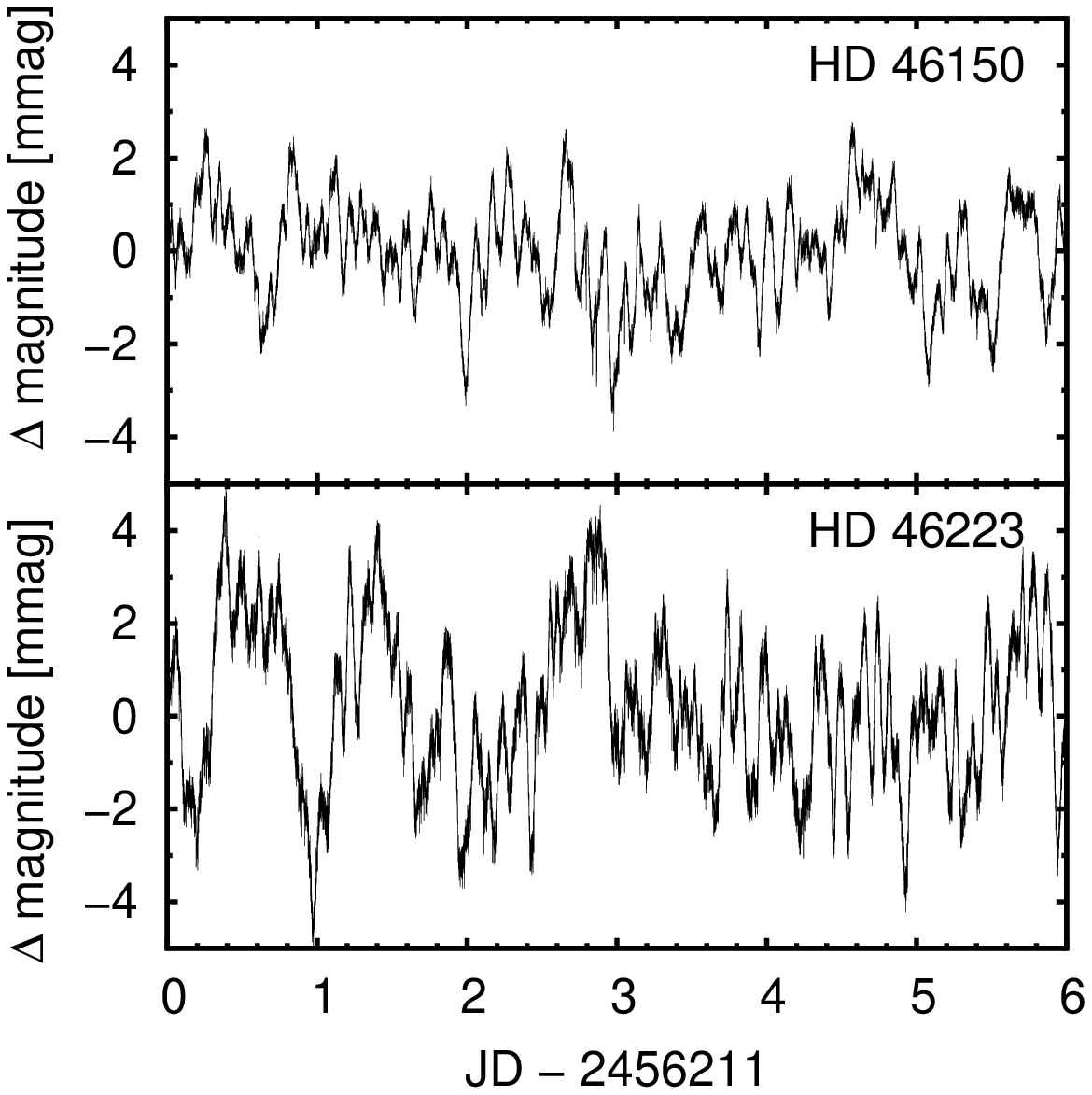}}
\resizebox{\hsize}{!}{\includegraphics{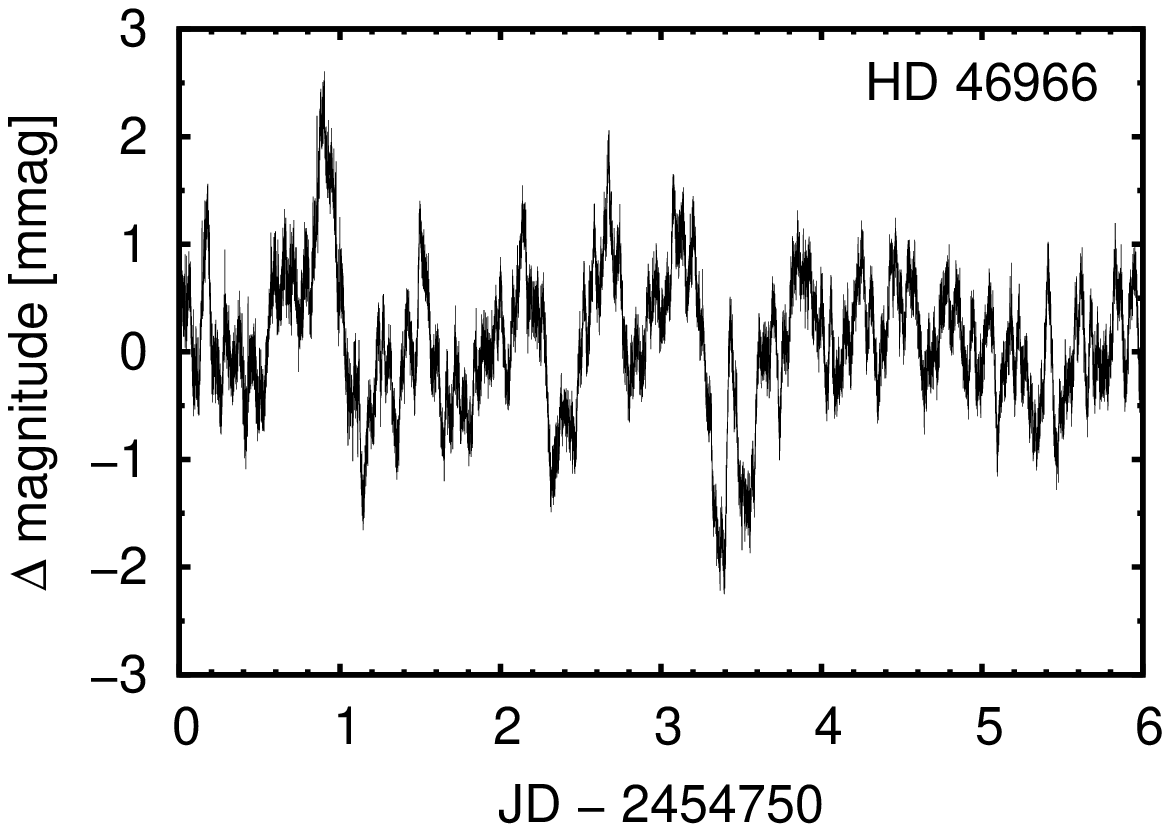}}
\caption{Portion of the light curve of O stars derived using the CoRoT
satellite. Difference between the actual and mean magnitudes is plotted.}
\label{ngc2244}
\end{figure}

The open cluster NGC 2244 observational run of the CoRoT satellite provides even
more promising results. The photometric study of the O9V star \object{HD 46202}
\citep{bricor} revealed multiple frequencies in the range $2-5\,\text{d}^{-1}$
and amplitude of the order of $0.01\,\text{mmag}$. This can be explained as a
result of wind blanketing due to multiple small-scale wind perturbations. O
stars \object{HD 46150}, \object{HD 46223}, and \object{HD 46966} show small
amplitude stochastic variations \citep[see also Fig.~\ref{ngc2244}]{blomcor}
that can be connected to the subphotospheric convection \citep{aro}. Comparing
with Fig.~\ref{d9ata_stormg_varms} follows that such variability can be caused
by a relatively large number of wind perturbations and wind blanketing. 

\begin{figure}[t]
\centering
\resizebox{\hsize}{!}{\includegraphics{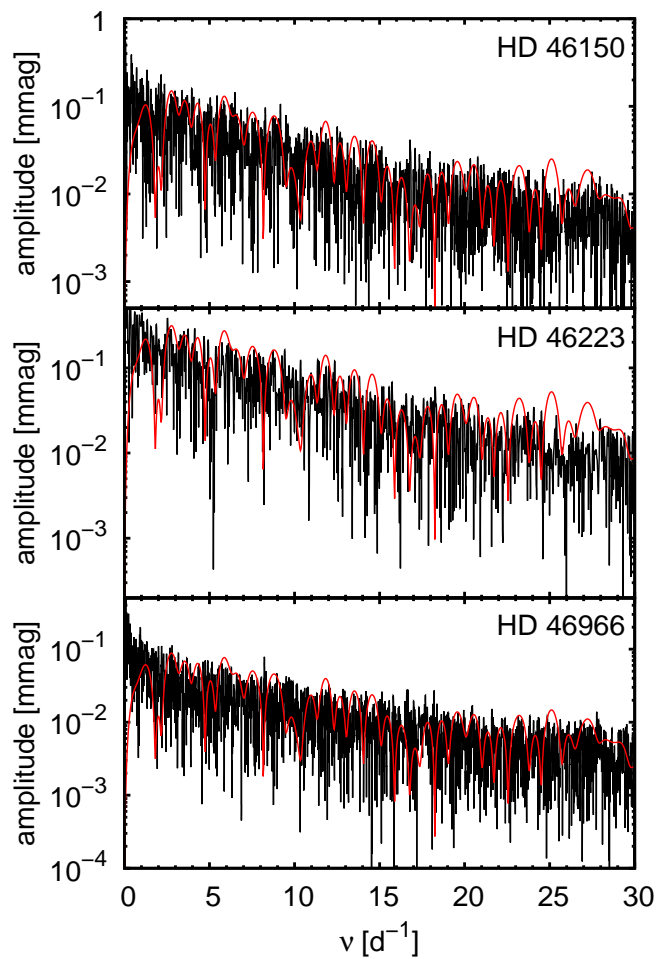}}
\caption{Comparison of the Fourier spectra of light curves given in
Fig.~\ref{ngc2244} (black curves) with Fourier spectrum of the theoretical light
curve for $N=1$ (red curves). The theoretical spectrum was scaled by a factor of
$10^b$ taken from Table~\ref{fittab} accounting for $N>1$. For the analysis we
used the CoRoT light curves secured between JD 2456203 -- 2456234 for HD~46150
and HD~46223 and between JD 2454748 -- 2454783 for HD~46966.}
\label{ngc2244f}
\end{figure}

This conclusion is further supported by the comparison of the Fourier spectra
(Fig.~\ref{ngc2244f}) and their fit parameters in Table~\ref{fittab}. The slope
parameters $a$ of the spectra based on predicted light curves are close to the
theoretical results. From Eq.~\eqref{bvarn} it follows that the observed light
variability is likely caused by a very large number of cones $N\sim10^2-10^4$,
which is consistent with results of numerical simulations \citep{sundsim}.
Numerical tests using the uncertainties of observed data showed that the Fourier
spectrum in Fig.~\ref{ngc2244f} is at least one order of magnitude higher than
that of observational noise. Therefore, the Fourier spectra in
Fig.~\ref{ngc2244f} should correspond to stellar variations.

\section{Discussion and conclusions}

We simulated the light variability of O stars due to variable wind blanketing
modulated by line-driven wind instability. We used the output from
hydrodynamical wind simulations and combined the derived time dependence of the
mass-loss rate with the dependence of optical flux on the wind mass-loss rate
derived from global wind models.

The resulting light curve has amplitude of the order of hundredths of magnitude
with a typical time scale of several hours. The amplitude is of the order of
millimagnitudes (and lower) when assuming that the wind consists of concentric
cones in which the wind behaves independently. Such variability is still
observable using high precision photometry.

We compared the derived light curve with light curve of $\epsilon$~Ori obtained
using the BRITE satellite and with light curves of O stars from open cluster
NGC~2244 obtained by CoRoT satellite. The observed light curves show stochastic
variability with amplitude and time scale comparable to that derived from
simulations. The light curves of NGC~2244 stars yield Fourier spectra that have
the same shape as the Fourier spectra of the predicted light curves and the
amplitudes that imply presence of large number ($N\sim10^2-10^4$) of independent
surface patches. The shape of the Fourier spectrum is slightly different for
$\epsilon$~Ori, but this may be connected with a narrow range of frequencies
available for analysis.

The variability due to the wind blanketing is another source of the light
variability in O stars, which contributes to the light variability due to the
pulsation and convection \citep{aro,buy}. Consequently, it may be problematic to
distinguish the wind blanketing variability from other effects. Ultraviolet
observations may solve this problem and test the proposed variability mechanism,
because the light variability in the ultraviolet (e.g. around 1300\,\AA) should
be in the anti-phase with visual variability \citep{magvar}. Moreover, there may
be stars that are variable only because of the wind blanketing, because only
perturbations of surface density and velocity (and no variations of the
frequency-integrated radiative flux) are required to trigger this type of
variability.

The proposed mechanism requires relatively large mass-loss rate variations.
Consequently, we do not expect strong light variations connected with
corotating  interaction regions \citep{crow,inver,lobl}. These regions have
been invoked to explain the discrete absorption components commonly observed in
the ultraviolet lines of hot stars. The discrete absorption components are
connected with velocity plateaus in the wind and require order of
magnitude lower mass-loss rate variations than those invoked by wind
instabilities \citep{duo}. Therefore, from Eq.~\eqref{hpt} we expect up to
millimagnitude variations due to corotating  interaction regions.

The light variability due to the wind blanketing is another piece of
puzzle to the general picture of variability in O stars. The subsurface
convective motions trigger surface light variations connected with pulsations and
variable wind blanketing leading to perturbations that disseminate in the wind
initiating the line-driven wind instability.

\begin{acknowledgements}
JK acknowledges support by grant GA\,\v{C}R  16-01116S. Based on data
collected by the BRITE Constellation satellite mission, designed, built,
launched, operated and supported by the Austrian Research Promotion Agency
(FFG), the University of Vienna, the Technical University of Graz, the Canadian
Space Agency (CSA), the University of Toronto Institute for Aerospace Studies
(UTIAS), the Foundation for Polish Science \& Technology (FNiTP MNiSW), and
National Science Centre (NCN). The CoRoT space mission, launched on 2006
December 27, was developed and operated by the CNES, with participation of
the Science Programs of ESA, ESA's RSSD, Austria, Belgium, Brazil, Germany and
Spain.
\end{acknowledgements}

\end{document}